\documentclass[12pt]{article}
\begin{document}

\def\ds{\displaystyle}
\def\beq{\begin{equation}}
\def\eeq{\end{equation}}
\def\bea{\begin{eqnarray}}
\def\eea{\end{eqnarray}}
\def\beeq{\begin{eqnarray}}
\def\eeeq{\end{eqnarray}}
\def\ve{\vert}
\def\vel{\left|}
\def\ver{\right|}
\def\nnb{\nonumber}
\def\ga{\left(}
\def\dr{\right)}
\def\aga{\left\{}
\def\adr{\right\}}
\def\lla{\left<}
\def\rra{\right>}
\def\rar{\rightarrow}
\def\nnb{\nonumber}
\def\la{\langle}
\def\ra{\rangle}
\def\ba{\begin{array}}
\def\ea{\end{array}}
\def\tr{\mbox{Tr}}
\def\ssp{{\Sigma^{*+}}}
\def\sso{{\Sigma^{*0}}}
\def\ssm{{\Sigma^{*-}}}
\def\xis0{{\Xi^{*0}}}
\def\xism{{\Xi^{*-}}}
\def\qs{\la \bar s s \ra}
\def\qu{\la \bar u u \ra}
\def\qd{\la \bar d d \ra}
\def\qq{\la \bar q q \ra}
\def\gGgG{\la g^2 G^2 \ra}
\def\q{\gamma_5 \not\!q}
\def\x{\gamma_5 \not\!x}
\def\g5{\gamma_5}
\def\sb{S_Q^{cf}}
\def\sd{S_d^{be}}
\def\su{S_u^{ad}}
\def\ss{S_s^{??}}
\def\sbp{{S}_Q^{'cf}}
\def\sdp{{S}_d^{'be}}
\def\sup{{S}_u^{'ad}}
\def\ssp{{S}_s^{'??}}

\def\sig{\sigma_{\mu \nu} \gamma_5 p^\mu q^\nu}
\def\fo{f_0(\frac{s_0}{M^2})}
\def\ffi{f_1(\frac{s_0}{M^2})}
\def\fii{f_2(\frac{s_0}{M^2})}
\def\O{{\cal O}}
\def\sl{{\Sigma^0 \Lambda}}
\def\es{\!\!\! &=& \!\!\!}
\def\ap{\!\!\! &\approx& \!\!\!}
\def\ar{&+& \!\!\!}
\def\ek{&-& \!\!\!}
\def\kek{\!\!\!&-& \!\!\!}
\def\cp{&\times& \!\!\!}
\def\se{\!\!\! &\simeq& \!\!\!}
\def\eqv{&\equiv& \!\!\!}
\def\kpm{&\pm& \!\!\!}
\def\kmp{&\mp& \!\!\!}
\def\mcdot{\!\cdot\!}

% .........................................................

\def\simlt{\stackrel{<}{{}_\sim}}
\def\simgt{\stackrel{>}{{}_\sim}}

% .........................................................

\title{
         {\Large
                 {\bf
Scalar form factor of the nucleon and nucleon--scalar
meson coupling constant in QCD
                 }
         }
      }

\author{\vspace{1cm}\\
{\small T. M. Aliev \thanks
{e-mail: taliev@metu.edu.tr}~\footnote{permanent address:Institute
of Physics,Baku,Azerbaijan}\,\,,
M. Savc{\i} \thanks
{e-mail: savci@metu.edu.tr}} \\
{\small Physics Department, Middle East Technical University,
06531 Ankara, Turkey} }

\date{}

\begin{titlepage}
\maketitle
\thispagestyle{empty}

\begin{abstract}
Scalar form factor of the nucleon is calculated in the framework of light
cone QCD sum rules, using the most general form of the baryon current. Using
the result on scalar form factor of the nucleon, the nucleon--scalar
$\sigma$ and $a_0$ meson coupling
constants are estimated. Our results on these couplings are in good agreement with
the prediction of the external field QCD sum rules method.  
\end{abstract}

%\vspace{1cm}
~~~PACS number(s): 11.55.Hx, 13.40.Em, 14.20.Jn
\end{titlepage}

\section{Introduction}

In spite of the great success of QCD theory of strong interactions,
the quark structure of scalar mesons is still an open question and widely
discussed in literature (see \cite{R8101} for a review). There are many
scenarios for the scalar meson structures like two--quark $\bar{q}q$,
four--quark $\bar{q}q\bar{q}q$ states, meson--meson bound states and even a
scalar glueball. It is very likely that the scalar meson structure is not
made of such simple structures, but are superpositions of these contents.
Therefore different scenarios can give quite different predictions on the
production and decays of scalar mesons which can be tested in experiments. 
One efficient way to make progress about the nature scalar bosons is
to study their role in the various two--baryon reactions ($NN$, $\Lambda N$,
$\Sigma N$, $\Lambda \Lambda$, $\Sigma \Sigma)$.

Our aim in this paper is to calculate the scalar form factor of the proton
in the framework of light cone sum rules (LCSR) \cite{R8102,R8103} using the
most general form of the proton current and then relate it to the
nucleon--scalar $\sigma$ or $a_0$ mesons coupling constants. 
This method is an alternative
approach to the traditional sum rule \cite{R8104}. LCSR is based on the
operator product expansion (OPE) near the light cone, which is an expansion 
of the time ordered product over the twist rather than the dimension of the
operators. Main contribution of this approach comes from the operators
having lower twist. The main ingredients of the LCSR are the wave functions
of hadrons which define the matrix elements of non--local operators between
the vacuum and one hadron states.

This method applied to study the nucleon electromagnetic form factors
\cite{R8105}, and the weak $\Lambda_b \rar p \ell \bar{\nu}_\ell$ decay
\cite{R8106}. The higher twist wave functions of baryons have been obtained
recently in \cite{R8107}.

The plan of this article is as follows: In section 2 we derive the sum rule 
for the proton form factor due to the scalar current. Section 3 is devoted 
to the numerical analysis and discussion.

\section{Sum rules for the proton form factor due to the scalar current}

In the present section we derive sum rules for the scalar form factor of the
proton using the most general form of nucleon interpolating current. For
this aim we start with the following correlation function:
\bea
\label{e8101}
\Pi(p,q) = i \int d^4 x \, e^{i q x} \lla 0 \vel T\{\eta(0)
J(x) \} \ver N(p) \rra~,
\eea
where $J(x)= \bar{u}u+(-1)^I \bar{d}d$ is the interpolating current for
the scalar $\sigma$ ($I=0$) and $a_0$ ($I=1$) mesons, respectively, and
$\eta$ is a suitable interpolating current with nucleon quantum number. Few
words about the interpolating current of the nucleon are in order. There
exist the following currents with the isotopic spin $1/2$, without
derivative terms (see \cite{R8108}):
\bea
\label{e8102}
\eta_1(x) \es 2 \varepsilon^{abc} \sum_{\ell=1}^2 \ga u^{Ta}(x) C A_1^\ell
d^b(x) \dr A_2^\ell u^c(x)~, \\
\label{e8103}
\eta_2(x) \es \frac{2}{3} \varepsilon^{abc} \Big[ \ga u^{Ta}(x) C \rlap/z
u^b(x) \dr \gamma_5 \rlap/z d^c(x) -
\ga u^{Ta}(x) C \rlap/z d^b(x) \dr \gamma_5 \rlap/z u^c(x)\Big]~,
\eea
where $A_1^1 = I$, $A_2^1=\gamma_5$, $A_1^2=\gamma_5$, $A_2^2= \beta$, and
in $C$ is the charge conjugation operator, $a,b,c$ are the color indices, and
$z$ in Eq. (\ref{e8103}) is a light--like vector with $z^2=0$. 
In Eq. (\ref{e8102}), $\beta=-1$
corresponds to the Ioffe current \cite{R8109}.
Note that in \cite{R8105} $\eta_2(x)$ is modified as
\bea
\eta_3(x) = \varepsilon^{abc} \Big[ \ga u^{Ta}(x) C \rlap/z   
u^b(x) \dr \gamma_5 \rlap/z d^c(x)\Big]~. \nnb
\eea
The nucleon scalar form factor which is obtained using the current
$\eta_3(x)$ in the light cone QCD sum rule approach is studied in \cite{R8110}. 
However, the
current $\eta_3(x)$ couples also to the states with isospin $3/2$, and these
contributions should be eliminated from the final result, which is not done
in \cite{R8110}. Moreover, terms that are proportional to $\rlap/z\rlap/x$
and $\rlap/x\rlap/z$ in the coordinate space which give contribution to the
considered structure in \cite{R8111}, are also neglected.
In the present work $\eta_1$ is chosen as the general form of the interpolating 
current.

Let us firstly calculate the hadronic part of the correlator function. By
inserting the complete set of states between the currents in Eq.
(\ref{e8101}) with quantum numbers of the corresponding nucleon, we get the
following hadronic representation of the correlator
\bea
\label{e8104}
\Pi(p,q) \es \frac{\lla 0 \vel \eta \ver N(p^\prime) \rra
\lla N(p^\prime) \vel J \ver N(p) \rra}
{m_N^2-p^{\prime 2}} + 
\sum_{h} \frac{\lla 0 \vel \eta \ver h(p^\prime) \rra
\lla h(p^\prime) \vel J \ver N (p) \rra}
{m_{h}^2-p^{\prime 2}}~,
\eea
where $p^\prime = p - q$, and $q$ is the momentum carried by the scalar
current. The second term in
Eq. (\ref{e8104}) takes into account higher states and continuum contributions
and hadrons $h$ form a complete set of baryons having the same quantum number as
the ground state baryon $B$ (here $B$ represents
$p$ and $n$). The matrix elements in Eq. (\ref{e8104}) are defined as
\bea
\label{e8105}
\lla 0 \vel \eta \ver N(p^\prime) \rra \es \lambda_N u_N(p^\prime)~,\\
\label{e8106}
\lla N(p^\prime) \vel J \ver N(p) \rra \es g(Q^2) \bar{u}(p^\prime) u(p)~.
\eea
Using Eqs. (\ref{e8104})--(\ref{e8106}) and performing summation over
nucleon spin, for the phenomenological part of the correlator we obtain
\bea
\label{e8107}
\Pi(p,q)  = \frac{\lambda_N g(Q^2) (\not\!p^\prime + m_N) u(p)}
{m_N^2-(p-q)^2} + \cdots~,
\eea
where $\cdots$ represents contributions from the higher states and the
continuum. It follows from Eq. (\ref{e8107}) that the correlator function
contains three different structures, namely structures $\sim \not\!p$,
$\sim \not\!q$ and structure proportional to the unit operator $I$. 
In further analysis we will choose the structure $\sim \not\!p$.

Here we would like to make the following remark: The nucleon--scalar meson
$S$ coupling
constant $g_{NNS}$ can be related to the scalar form factor $g(Q^2)$,
and for this aim we define the following coupling constant
\bea
\label{e8108}
\lla N(p^\prime) S \ve N(p) \rra \equiv g_{NNS} \bar{u} (p^\prime)
u(p)~.
\eea
It follows then from Eq. (\ref{e8101}) that
\bea
\label{e8109}
\Pi(p,q)  = \frac{\lla 0 \vel J \ver S \rra
                  \lla N(p^\prime) S \ve N(p) \rra
                  \lla 0 \vel \eta \ver N(p^\prime) \rra }
                 {(m_N^2-p^{\prime 2}) (m_S^2 - q^2) }~.
\eea
The matrix element of the scalar current $J$ between the vacuum and the
scalar meson state is defined as
\bea
\label{e8110}
\lla 0 \vel J \ver S \rra = \lambda_S~.
\eea
Using Eqs. (\ref{e8108})--(\ref{e8110}) and comparing with Eq.
(\ref{e8107}), we get the following relation between nucleon scalar form
factor and $g_{NNS}$
\bea
\label{e8111}
g(Q^2) = \frac{g_{NNS} \lambda_S}{m_S^2+Q^2}~.
\eea   
At large Euclidean momenta $p^{\prime 2} = (p-q)^2$ and $q^2=-Q^2$, 
the QCD part of the correlator can be calculated from the explicit form of 
interpolating currents $J(x)$ and $\eta(0)$, from which we obtain
\bea
\label{e8112}
\Pi \es \frac{1}{2} \int d^4x e^{iqx} \sum_{\ell=1}^2 \Bigg\{
\ga C A_1^\ell\dr_{\alpha\gamma} \Big[ A_2^\ell S(-x) \Big]_{\rho\beta}
4 \epsilon^{abc} \lla 0 \vel T\Big\{u_\alpha^a(0) u_\beta^b(x)
d_\gamma^c(0)\Big\} \ver N \rra \nnb \\
\ar \ga A_2^\ell \dr_{\rho\alpha} \Big[ \ga C A_1^\ell\dr^T S(-x)
\Big]_{\gamma\beta} 4 \epsilon^{abc} \lla 0 \vel T\Big\{u_\alpha^a(0)
u_\beta^b(x) d_\gamma^c(0)\Big\} \ver N \rra \nnb \\
\ar \ga A_2^\ell \dr_{\rho\beta} \Big[ C A_1^\ell S(-x)
\Big]_{\alpha\gamma} 4 \epsilon^{abc} \lla 0 \vel T\Big\{u_\alpha^a(0)
u_\beta^b(0) \Big\} d_\gamma^c(x) \ver N \rra \Bigg\}~, 
\eea
where $S(x)$ is the light quark propagator. It follows from Eq.
(\ref{e8112}) that for calculation of the correlator from QCD side we need
to know the explicit form light quark propagator and the nucleon
distribution amplitudes (DAs). The full propagator of the massless quark is 
\bea
\label{e8113}
S(x) = \frac{i \not\!x}{2 \pi^2 x^4} - \frac{\la q \bar{q} \ra}{12} \Bigg(
1+\frac{m_0^2 x^2}{16} \Bigg)
- i g_s \int_0^1 dv \Bigg[ \frac{\not\!x}{16 \pi^2 x^4} G_{\mu\nu}
\sigma^{\mu\nu} - v x^\mu G_{\mu\nu} \gamma^\nu \frac{i}{4 \pi^2 x^2}
\Bigg]~.
\eea
The contribution $\sim G_{\mu\nu}$ can give rise to four and five particle
nucleon DAs. These DAs are usually
small \cite{R8112, R8113} and can be neglected.

The second term in Eq. (\ref{e8113}) does not give any contribution to the
sum rule after the Borel transformation. So, only the first term is relevant
for our calculations.

As has already been noted, in order to calculate the non--perturbative
contributions to the theoretical part of the correlator function one needs
to know the matrix element of $\epsilon^{abc} \la 0 \ve u_\alpha^a(a_1x) 
u_\beta^b(a_2x) d_\gamma^c(a_3x) \ve N(p) \ra$ between the proton and
vacuum states. This matrix element of the non--local operator is parametrized
in terms of the nucleon DAs
\cite{R8107,R8113,R8114,R8115,R8116,R8117}. Their explicit forms are given
in the Appendix--A.

Taking into account the three valence quark light-cone DAs
up to twist-6 and performing integration over $x$
in the coordinate space, we finally obtain the following expression for the
correlator function:
\bea
\label{e8114}
\Pi \es \frac{1}{4} m_N \int_0^1 dt_2 \int_0^{1-t_2} dt_1
\frac{\not\!q-\not\!p t_2}{(q-p t_2)^2} 
\Big\{ 2(1+\beta)
\widetilde{\cal V}_2(t_i) + 2[3 \mp 1 +(3\pm 1)\beta] V_3(t_i) \nnb \\
\kpm 2(1+\beta) P_1(t_i) - 2 [2 \pm 1 - (2\mp 1) \beta] S_1(t_i) + 
2(1+\beta) \widetilde{\cal A}_2(t_i) \nnb \\
\ar 2[ 3\pm 1 + (3 \mp 1) \beta] {\cal A}_3(t_i)
+ [2 \mp 1 - (2 \pm 1) \beta] \widetilde{\cal T}_2(t_i) + 12 (1-\beta) 
T_7(t_i) \nnb \\
\ar [6 \mp 1 - (6 \pm 1) \beta] \widetilde{\cal T}_4(t_i)
\Big\} \nnb \\
\ek \frac{m_N^3}{2} \int_0^1 d\tau \int_1^\tau d\lambda \int_1^\lambda dt_2
\int_0^{1-t_2} dt_1 \frac{\not\!q-\not\!p \tau}{(q-p \tau)^4}
\Big\{ \pm (1-\beta) \widetilde{\cal V}_6(t_i) \nnb \\
\kmp (1-\beta) \widetilde{\cal A}_6(t_i)
+ 3 [ 2 \mp 1 - (2 \pm 1) \beta] \widetilde{\cal T}_8(t_i) \Big\} \nnb \\
\ek \frac{m_N}{2} \int_0^1 d\lambda \int_1^\lambda dt_2 \int_0^{1-t_2}
dt_1  \frac{\not\!p}{(q-p \lambda)^2}
\Big\{
[ 2 \mp 1 + (2 \pm 1) \beta] \widetilde{\cal V}_2(t_i) \nnb \\
\ar [ 2 \pm 1 + (2 \mp 1) \beta] \widetilde{\cal A}_2(t_i)
+ [ 2 \mp 1 - (2 \pm 1) \beta] \widetilde{\cal T}_2(t_i) \nnb \\
\kpm 2 (1+\beta) \widetilde{\cal T}_4(t_i)
\Big\}
\eea
where any function $F(t_i)$ means $F(t_i)=F(t_1,t_2,1-t_1,t_2)$, and
functions with $\sim$ are defined as
\bea
\widetilde{\cal V}_2(t_i) \es V_1(t_i) - V_2(t_i) - V_3(t_i)~, \nnb \\
\widetilde{\cal A}_2(t_i) \es - A_1(t_i) + A_2(t_i) - A_3(t_i)~, \nnb \\
\widetilde{\cal A}_6(t_i) \es A_1(t_i) - A_2(t_i) + A_3(t_i) + 
A_4(t_i) - A_5(t_i) + A_6(t_i)~, \nnb \\
\widetilde{\cal T}_2(t_i) \es T_1(t_i) + T_2(t_i) - 2 T_3(t_i)~, \nnb \\
\widetilde{\cal T}_4(t_i) \es T_1(t_i) - T_2(t_i) - 2 T_7(t_i)~, \nnb \\
\widetilde{\cal T}_6(t_i) \es 2 T_2(t_i) - 2 T_3(t_i) - 2 T_4(t_i) + 
2 T_5(t_i) + 2 T_7(t_i) + 2 T_8(t_i)~, \nnb \\
\widetilde{\cal T}_8(t_i) \es - T_1(t_i) + T_2(t_i) + T_5(t_i) - 
T_6(t_i) + 2 T_7(t_i) + 2 T_8(t_i)~. \nnb
\eea
Explicit expressions of $A_\alpha(t_i)$, $T_\alpha(t_i)$ and
$V_\alpha(t_i)$ can be found in \cite{R8117}.
Note also that in deriving Eq. (\ref{e8114}) we omit the terms proportional with
the unit operators which do not give contribution to the structure
$\sim\not\!p$. Upper (lower) sign in Eq. (\ref{e8114}) corresponds to the
scalar form factor due to the scalar current with the quantum numbers of
$\sigma$ ($a_0$).

In order to suppress the continuum and higher state contributions we need to
apply Borel transformation to the phenomenological and theoretical parts of
the correlation function with respect to the variable $(q-p)^2$.

For the theoretical part the Borel transformation and the continuum
subtraction can be done by using the following substitution rules
\bea
\label{e8115}
\int dx \frac{\rho(x)}{(q-xp)^2} &\rar& - 
\int dx \frac{\rho(x)}{x} e^{-s/M^2}~, \nnb \\ 
\int dx \frac{\rho(x)}{(q-xp)^4} &\rar& \frac{1}{M^2}
\int dx \frac{\rho(x)}{x^2} e^{-s/M^2}
+ \frac{\rho(x_0)}{Q^2+x_0^2 M^2}e^{-s_0/M^2}~,
\eea
where 
\bea
s \es (1-x) M^2 + \frac{1-x}{x} Q^2~, \nnb \\ \nnb \\
x_0 \es \frac{\sqrt{(Q^2+s_0-m_N^2)^2 + 4 m_N^2 Q^2} - (Q^2+s_0-m_N^2)}
{2 m_N62}~, \nnb
\eea
and $Q^2 = - q^2$.
Finally, we get the following sum rule for the scalar form factor.
\bea
\label{e8116}
g(Q^2) \es
\frac{e^{m_N^2/M^2}}{\lambda_N}\Bigg[
\frac{1}{4} m_N \int_{x_0}^1 dt_2 \int_0^{1-t_2} e^{-s/M^2}
\Big\{ 2(1+\beta)
\widetilde{\cal V}_2(t_i) + 2[3 \mp 1 +(3\pm 1)\beta] V_3(t_i) \nnb \\
\kpm 2(1+\beta) P_1(t_i) - 2 [2 \pm 1 - (2\mp 1) \beta] S_1(t_i) + 
2(1+\beta) \widetilde{\cal A}_2(t_i) \nnb \\
\ar 2[ 3\pm 1 + (3 \mp 1) \beta] {\cal A}_3(t_i)
+ [2 \mp 1 - (2 \pm 1) \beta] \widetilde{\cal T}_2(t_i) + 12 (1-\beta) 
T_7(t_i) \nnb \\
\ar [6 \mp 1 - (6 \pm 1) \beta] \widetilde{\cal T}_4(t_i)
\Big\} \nnb \\
\ar \frac{m_N^3}{2} \Bigg(\int_{x_0}^1 d\tau \int_1^\tau d\lambda \int_1^\lambda dt_2
\int_0^{1-t_2} dt_1 \frac{1}{\tau M^2} e^{-s/M^2}
\Big\{ \pm (1-\beta) \widetilde{\cal V}_6(t_i) \nnb \\
\kmp (1-\beta) \widetilde{\cal A}_6(t_i)
+ 3 [ 2 \mp 1 - (2 \pm 1) \beta] \widetilde{\cal T}_8(t_i) \Big\} \nnb \\
\ar \frac{x_0}{Q^2+x_0^2 m_N^2} \int_1^{x_0} d\lambda \int_1^\lambda dt_2
\int_0^{1-t_2} dt_1 \frac{1}{\tau M^2} e^{-s/M^2}
\Big\{ \pm (1-\beta) \widetilde{\cal V}_6(t_i) \nnb \\
\kmp (1-\beta) \widetilde{\cal A}_6(t_i)
+ 3 [ 2 \mp 1 - (2 \pm 1) \beta] \widetilde{\cal T}_8(t_i) \Big\} \Bigg) \nnb \\
\ar \frac{1}{2} m_N \int_{x_0}^1 \frac{dt_2}{t_2}\int_0^{1-t_2}
dt_1 e^{-s/M^2} \Big\{
[ 2 \mp 1 + (2 \pm 1) \beta] \widetilde{\cal V}_2(t_i) \nnb \\
\ar [ 2 \pm 1 + (2 \mp 1) \beta] \widetilde{\cal A}_2(t_i) 
+ [ 2 \mp 1 - (2 \pm 1) \beta] \widetilde{\cal T}_2(t_i) \nnb \\
\kpm 2 (1+\beta) \widetilde{\cal T}_4(t_i)
\Big\}
\Bigg]
\eea
It follows from Eq. (\ref{e8116}) that for determining the scalar form
factor, the residue $\lambda_N$ of the nucleon needs to be known. The
residue $\lambda_N$ is obtained from the mass sum rules for the baryons (see
for example \cite{R8118})
\bea
\label{e8117}
\lambda_N^2 e^{-m_N^2/M^2} \es \frac{M^6}{256 \pi^4} E_2(x) (5+2 \beta +
\beta^2) + \frac{\la \bar{u}u \ra}{6} \Big[ - 6 (1-\beta^2) \la \bar{d}d \ra +
(-1+\beta^2) \la \bar{u}u \ra \Big] \nnb \\
\ek \frac{m_0^2}{24 M^2}  \la \bar{u}u \ra \Big[- 12 (1-\beta^2) \la \bar{d}d
\ra + (-1+\beta^2) \la \bar{u}u \ra \Big]~,
\eea
where $x=s_0/M^2$, and 
\bea
E_n(x) = 1-e^x \sum_{k=0}^n \frac{x^k}{k!}~, \nnb
\eea 
which describes the subtraction of higher states and continuum
contributions.

\section{Numerical Analysis}
In this section we present the numerical analysis for the scalar form factor
$g(Q^2)$. It follows from Eq. (\ref{e8116}) that the main ingredients of LCSR
are the DAs, which involve non--perturbative parameters $f_N$, $\lambda_1$, 
$\lambda_2$, $f_1^u$, $f_1^d$, $f_2^d$, $A_1^u$, $V_1^d$ (see Appendix--B).

In the numerical calculations we use three different sets of the DAs (for
more details, see \cite{R8105} and \cite{R8117}).

1) QCD sum rules based DAs,

2) Asymptotic DAs,

3) A model for the nucleon DAs. The parameters in this case are chosen in
such a way that, the form of DAs describes very well the existing
experimental data on nucleon form factors.

The values of the above--mentioned non--perturbative parameters for these
cases at $\mu=1~GeV$ scale are given in \cite{R8105} and \cite{R8117}. Our
numerical calculations show that the values of the scalar form factors of
nucleon due to the scalar current in QCD based DAs case
are larger compared to that of the other two scenarios, and, for customary
reasons we present our numerical results only for this case.
  
In the problem at hand we have three auxiliary parameters, namely, Borel
mass square $M^2$, continuum threshold $s_0$ and $\beta$ in the
interpolating current of nucleon. Obviously, any physically measurable
must be independent of the auxiliary parameters. So, in the first hand we
have to find the appropriate regions of these parameters, in where 
the form factor is independent of them. The mass sum rules analysis shows
that optimal value for the continuum thresholds is in the region 
$s_0 (2 \div 2.5)~GeV^2$.

For this aim we consider the following three--step procedure. In the first
step we try to find the working region of $M^2$ where scalar form factor is
independent of $M^2$ at fixed values of $s_0$ and $\beta$. In Figs. (1),
(2), (3) and (4) (Figs. (5), (6), (7) and (8)) we present the dependence of
the scalar form factor induced by $I=0~(I=1)$scalar current on $M^2$ at
different fixed values of $Q^2$ and $\beta$ at $s_0=2.0~GeV^2$ and
$s_0=2.5~GeV^2$, respectively.

From these figures we see that, in both cases, 
the scalar form factor is practically independent of $M^2$ at different
values of $Q^2$, $\beta$ and $s_0$. To decide on an upper bound for $M^2$ we
require the contribution of continuum be less than the contribution of
continuum--subtracted sum rules and the lower bound can be determined by
requiring that the contribution of the highest power of $1/M^2$ to be less
than, say, $30\%$ of the higher powers of $M^2$. As a result of our analysis
we find that the region $1~GeV^2 \le M^2 \le 2.5~GeV^2$ satisfies both
conditions simultaneously. 

The parameter $\beta$ is an auxiliary parameter, and hence, the physical
quantities should be independent of it. For this reason we need to find the
working region of $\beta$ where the scalar form factor is independent of it.
In determining this region of $\beta$ two conditions should be satisfied:
Eq. (\ref{e8117}) and its logarithmic derivative with respect to
$1/M^2$ should be positive and independent of $\beta$. From an analysis of
Eq. (\ref{e8117}) it is found that $-0.5 < \beta < 0.5$ is unphysical
\cite{R8118}. It is further shown in \cite{R8118} that the working region
of $\beta$ where the second condition is satisfied is obtained for $-0.6
< \cos\theta < 0.3$, where $\tan\theta=\beta$, which yields $\beta < -1.3$
and $\beta > 3.3$.   

We perform calculations depicting the dependence of the scalar form factors
on $\cos\theta$ at fixed values of $s_0$ ($s_0=2.0~GeV^2$ and
$s_0=2.5~GeV^2$), $M^2$ ($M^2=1.0~GeV^2$ and $M^2=1.5~GeV^2$) at three fixed
values of $Q^2$ ($Q^2=3.0~GeV^2$, $Q^2=4.0~GeV^2$ and $Q^2=5~GeV^2$); and
confirmed that in the regions when $\beta < -1.3$ and $\beta > 3.3$ ,the
form factor is practically independent of $\beta$. So, the common region
where mass sum and form factor are independent of $\beta$ is the region
$-0.5 < \cos\theta < 0.3$, which corresponds to $\beta < -1.3$ and $\beta >
3.3$ and which we will use in further discussions. Note that, the Monte Carlo
analysis that is performed in \cite{R8119} predicts the optimal value of
$\beta$ to be $\beta=-1.2$. Our lower bound is slightly different compared
to this value.   

We perform the OPE in the light cone with large Euclidean $Q^2$ and $(p-q)^2$
momenta, where
the scalar from factor $g(Q^2)$ can reliably be determined at the range $Q^2
> 2~GeV^2$. In the region $Q^2 < 2~GeV^2$ OPE becomes questionable and for
this reason in determining $g(Q^2)$ we restrict ourselves to the region $Q^2
> 2~GeV^2$.

In Figs. (9), (10), (11) and (12) we present the dependence of $g(Q^2)$ 
induced by $I=0~(I=1)$ current on $Q^2$ at fixed
values of $\beta$ which lies in the above--mentioned region of $\beta$; at
$M^2= 1~GeV^2$ and $s_0=2~GeV^2$, $s_0=2.5~GeV^2$, respectively, for the
central values all parameters. We observe that the value of $g(Q^2)$ seems
to be insensitive to the choice of $s_0$. We further see from these figures
that for positive (negative) values $\beta$, the sign of the form factor
is positive (negative). Therefore determination of the sign of $g(Q^2)$ can
give valuable information about parameter $\beta$.

Having the obtained the $Q^2$ dependence of the form factor, we can
determine $g_{NNS}$. It follows from Eq. (\ref{e8110}) that
\bea
g_{NNS} = \frac{(m_S^2 + Q^2) g(Q^2)}{\lambda_S}~. \nnb
\eea

The residues $\lambda_\sigma$ and $\lambda_{a_0}$ of the scalar currents
are determined from
two--point QCD sum rules analysis in \cite{R8120,R8121}, which is predicted
to have the value $\lambda_\sigma = 0.2$ and $\lambda_{a_0} = 0.3$. 
Using the results of Figs. (3) and (4), for the scalar form factor 
$g_{NN\sigma}$ we get
\bea
\vel g_{NN\sigma} \ver = \left\{ \begin{array}{ll}
6 \pm 1 & \mbox{\rm for}~\beta= 5~,\\ \\
8 \pm 2 & \mbox{\rm for}~\beta=-5~;~~\beta=-3~,
\end{array} \right. \nnb
\eea

\bea
\vel g_{NNa_0} \ver = \left\{ \begin{array}{ll}
5.5 \pm 1.5 & \mbox{\rm for}~\beta= 5~,\\ \\
7 \pm 1 & \mbox{\rm for}~\beta=-5~;~~\beta=-3~.     
\end{array} \right. \nnb
\eea

For completeness we present the results for $g_{NNS}$ coupling for the Ioffe
current case, despite that $\beta=-1$ lies outside the working region of
$\beta$. Our results are
\bea
\vel g_{NN\sigma}\ver \es 12 \pm 2 ~, \nnb \\
\vel g_{NNa_0}\ver    \es 11 \pm 2 \nnb~.
\eea   

Note that his coupling constant in framework of the external--field QCD sum
rules method using the Ioffe current is evaluated in \cite{R8122}, and it is 
obtained that $g_{NN\sigma}=14.4 \pm 3.7$. This same coupling constant is 
also shown to have the value $g_{NN\sigma}=16.9$ \cite{R8101} from a fit to 
the $NN$ scattering data.

A comparison of our result for $\beta=-1$ shows that, within the error 
limits, there is a good 
agreement between our result and the above--mentioned one.

In conclusion, we have calculated the nucleon scalar form factor in the
framework of light cone QCD sum rules, using the most general form of the
baryon current. Using the obtained result on scalar form factor, we estimate
$g_{NNS}$. Our result in regard to the coupling
constant $g_{NN\sigma}$ is in a good agreement with the prediction of the 
external field QCD sum rules method for the Ioffe current. Measurement of
the nucleon scalar meson coupling constant can give valuable information
about the quark structure of scalar meson as well as about the value of
parameter $\beta$.

\section*{Acknowledgments}

One of the authors (T. M. A) is grateful to T\"{U}B\.{I}TAK for partially
support of this work under the project 105T131.

\newpage

\appendix  
\renewcommand{\theequation}{\Alph{section}.\arabic{equation}} 

\section*{Appendices}  
\section{Nucleon distribution amplitudes}
\setcounter{equation}{0}
In this appendix we present the three--quark distribution DAs from
twist--3 to twist--6 (for more details, see \cite{R8107} and \cite{R8117}).

The matrix element of the non--local operator $\epsilon^{abc} \la 0 \ve
u_\alpha^a(a_1x) u_\beta^b(a_2x) d_\gamma^c(a_3x) \ve N(p) \ra$ between 
the proton and vacuum states is parametrized in terms of the nucleon 
DAs as follows:
\bea
\label{e81A1}
&&4 \epsilon^{abc} \lla 0 \vel u_\alpha^a(a_1 x) u_\beta^b(a_2x)
d_\gamma^c(a_3x)\ver P\rra =
{\cal S}_1 m_N C_{\alpha\beta}(\gamma_5 N)_{\gamma} +
{\cal S}_2 m_N^2 C_{\alpha\beta}(\not\!x \gamma_5N)_{\gamma} \nnb \\
\ar {\cal P}_1 m_N (\gamma_5 C)_{\alpha\beta} N_{\gamma} +
{\cal P}_2 m_N^2 (\gamma_5 C)_{\alpha\beta}(\not\!x N)_{\gamma} +
\ga {\cal V}_1 + \frac{x^2 m_N^2}{4} {\cal V}_1^M \dr 
(\not\!P C)_{\alpha\beta} (\gamma_5 N)_{\gamma} \nnb \\
\ar {\cal V}_2 m_N (\not\!P C)_{\alpha\beta}(\not\!x \gamma_5 N)_{\gamma} +
{\cal V}_3 m_N(\gamma_\mu C)_{\alpha\beta}(\gamma^\mu \gamma_5 N)_{\gamma} +
{\cal V}_4 m_N^2(\not\!x C)_{\alpha\beta}(\gamma_5 N)_{\gamma} \nnb \\
\ar {\cal V}_5 m_N^2 (\gamma_\mu C)_{\alpha\beta}
(i\sigma^{\mu\nu} x_\nu \gamma_5 N)_{\gamma} +
{\cal V}_6 m_N^3 (\not\!x C)_{\alpha\beta}(\not\!x \gamma_5 N)_{\gamma} \nnb \\
\ar \ga {\cal A}_1 + \frac{x^2 m_N^2}{4} {\cal A}_1^M \dr
(\not\!P \gamma_5 C)_{\alpha\beta} N_{\gamma} +
{\cal A}_2 m_N (\not\!P \gamma_5 C)_{\alpha\beta}(\not\!x N)_{\gamma} +
{\cal A}_3 m_N (\gamma_\mu \gamma_5 C)_{\alpha\beta}(\gamma^\mu N)_{\gamma} \nnb \\
\ar {\cal A}_4 m_N^2 (\not\!x \gamma_5 C)_{\alpha\beta} N_{\gamma} +
{\cal A}_5 m_N^2 (\gamma_\mu \gamma_5 C)_{\alpha\beta}
(i\sigma^{\mu\nu} x_\nu N)_{\gamma} +
{\cal A}_6 m_N^3 (\not\!x \gamma_5 C)_{\alpha\beta}(\not\!x
N)_{\gamma} \nnb \\ 
\ar \ga {\cal T}_1 + \frac{x^2 m_N^2}{4} {\cal T}_1^M \dr 
(P^\nu i\sigma_{\mu\nu} C)_{\alpha\beta}(\gamma^\mu \gamma_5 N)_{\gamma} +
{\cal T}_2 m_N (x^\mu P^\nu i\sigma_{\mu\nu} C)_{\alpha\beta}(\gamma_5
N)_{\gamma}\nnb \\
\ar {\cal T}_3 m_N (\sigma_{\mu\nu} C)_{\alpha\beta}(\sigma^{\mu\nu} \gamma_5
N)_{\gamma} + {\cal T}_4 m_N (P^\nu \sigma_{\mu\nu} C)_{\alpha\beta}
(\sigma^{\mu\rho} x_\rho \gamma_5 N)_{\gamma} \nnb \\ 
\ar {\cal T}_5 m_N^2 (x^\nu i\sigma_{\mu\nu} C)_{\alpha\beta}
(\gamma^\mu\gamma_5 N)_{\gamma} + {\cal T}_6 m_N^2 (x^\mu P^\nu
i\sigma_{\mu\nu} C)_{\alpha\beta}(\not\!x \gamma_5 N)_{\gamma}\nnb \\
\ar {\cal T}_7 m_N^2 (\sigma_{\mu\nu} C)_{\alpha\beta}(\sigma^{\mu\nu} 
\not\!x \gamma_5 N)_{\gamma} + {\cal T}_8 m_N^3 
(x^\nu \sigma_{\mu\nu} C)_{\alpha\beta}
(\sigma^{\mu\rho}x_\rho\gamma_5 N)_{\gamma}~.
\eea
The calligraphic DAs do not have definite twist and can
be related to the one with definite twist as:
\bea
\begin{array}{ll}
{\cal S}_1 = S_1~,& (2 P \mcdot x) \, {\cal S}_2 = S_1 - S_2~, \nnb \\
{\cal P}_1 = P_1~,& (2 P \mcdot x) \, {\cal P}_2 = P_2 - P_1~, \nnb \\
{\cal V}_1 = V_1~,& (2 P \mcdot x) \, {\cal V}_2 = V_1 - V_2 - V_3~, \nnb \\ 
2{\cal V}_3 = V_3~,& (4 P \mcdot x) \, {\cal V}_4 = 
- 2 V_1 + V_3 + V_4 + 2 V_5~,\nnb \\ 
(4 P\mcdot x) \, {\cal V}_5 = V_4 - V_3~,& 
(2 P \mcdot x)^2 \, {\cal V}_6 = - V_1 + V_2 + V_3 + V_4 + V_5 - V_6~, \nnb \\
{\cal A}_1 = A_1~,& (2 P \mcdot x) \, {\cal A}_2 = - A_1 + A_2 - A_3~, \nnb \\ 
2 {\cal A}_3 = A_3~,& 
(4 P \mcdot x) \, {\cal A}_4 = - 2 A_1 - A_3 - A_4 + 2 A_5~, \nnb \\
(4 P \mcdot x) \, {\cal A}_5 = A_3 - A_4~,& 
(2 P \mcdot x)^2 \, {\cal A}_6 = A_1 - A_2 + A_3 + A_4 - A_5 + A_6~, \nnb \\
{\cal T}_1 = T_1~, & (2 P \mcdot x) \, {\cal T}_2 = T_1 + T_2 - 2T_3~, \nnb \\
2 {\cal T}_3 = T_7~,& (2 P \mcdot x) \, {\cal T}_4 = T_1 - T_2 - 2 T_7~, \nnb \\
(2 P\mcdot x) \, {\cal T}_5 = - T_1 + T_5 + 2 T_8~,&
(2 P \mcdot x)^2 \, {\cal T}_6 = 2 T_2 - 2 T_3 - 2 T_4 + 2 T_5 + 2 T_7 + 2 T_8~, \nnb \\
(4P \mcdot x) \, {\cal T}_7 = T_7 - T_8~, &
(2 P\mcdot x)^2 \, {\cal T}_8 = - T_1 + T_2 + T_5 - T_6 + 2 T_7 + 2 T_8~. \nnb
\end{array}
\eea
These DAs can be written as
\bea
\label{e81A2}
F(a_i \, p\cdot x) = \int dx_1 \,dx_2 \,dx_3 \, \delta(x_1+x_2+x_3-1) \,
e^{-ip \cdot x\Sigma_i x_i a_i}\,F(x_i)~.
\eea

These light cone DAs are scale dependent and can be
expanded next--to--leading spin accuracy with the conformal operators whose
explicit expressions are,
\bea
\label{e81A3}
V_1(x_i,\mu) \es 120 x_1 x_2 x_3[\phi_3^0(\mu) + 
\phi_3^+(\mu)(1 - 3 x_3)]~, \nnb \\
V_2(x_i,\mu) \es 24 x_1 x_2 [\phi_4^0(\mu) + \phi_4^+(\mu)
(1 - 5 x_3)]~, \nnb \\
V_3(x_i,\mu) \es 12 x_3 \Big\{ \psi_4^0(\mu)(1 - x_3) + 
\psi_4^-(\mu) [x_1^2 + x_2^2 - x_3 (1 - x_3)] \nnb \\
\ar \psi_4^+(\mu)(1 - x_3 - 10 x_1 x_2) \Big\}~, \nnb \\
V_4(x_i,\mu) \es 3 \Big\{ \psi_5^0(\mu) (1 - x_3) + 
\psi_5^-(\mu) [2 x_1 x_2 - x_3 (1 - x_3)] \nnb \\
\ar \psi_5^+(\mu) [1 - x_3 - 2 (x_1^2 + x_2^2)] \Big\}~, \nnb \\
V_5(x_i,\mu) \es 6 x_3 [\phi_5^0(\mu) + \phi_5^+(\mu) (1 - 2 x_3)]~, \nnb \\
V_6(x_i,\mu) \es 2 [\phi_6^0(\mu) + \phi_6^+(\mu) (1 - 3 x_3)]~, \nnb \\
A_1(x_i,\mu) \es 120 x_1 x_2 x_3 \phi_3^-(\mu) (x_2 - x_1)~, \nnb \\
A_2(x_i,\mu) \es 24 x_1 x_2 \phi_4^-(\mu) (x_2 - x_1)~, \nnb \\
A_3(x_i,\mu) \es 12 x_3 (x_2 - x_1) \Big\{ [\psi_4^0(\mu) + 
\psi_4^+(\mu) ] + \psi_4^-(\mu) (1 - 2 x_3) \Big\}~, \nnb \\
A_4(x_i,\mu) \es 3(x_2-x_1) \Big\{ - \psi_5^0(\mu) + \psi_5^-(\mu) x_3
+ \psi_5^+(\mu) (1 - 2 x_3) \Big\}~, \nnb \\
A_5(x_i,\mu) \es 6 x_3 (x_2 - x_1) \phi_5^-(\mu)~, \nnb \\
A_6(x_i,\mu) \es 2 (x_2 - x_1) \phi_6^-(\mu)~, \nnb \\
T_1(x_i,\mu) \es 120 x_1 x_2 x_3 [ \phi_3^0(\mu) + \frac{1}{2}
(\phi_3^- - \phi_3^+)(\mu) (1 - 3 x_3) ]~, \nnb \\
T_2(x_i,\mu) \es 24 x_1 x_2 [ \xi_4^0(\mu) + 
\xi_4^+(\mu) (1 - 5 x_3) ]~, \nnb \\
T_3(x_i,\mu) \es 6 x_3 \Big\{ (\xi_4^0 + \phi_4^0 + \psi_4^0)(\mu) (1 - x_3) +
(\xi_4^- + \phi_4^- - \psi_4^-)(\mu) [x_1^2 + x_2^2 - x_3 (1 - x_3)] \nnb \\
\ar (\xi_4^+ + \phi_4^+ + \psi_4^+)(\mu) (1 - x_3 - 10 x_1 x_2) \Big\}~, \nnb \\
T_4(x_i,\mu) \es \frac{3}{2} \Big\{ (\xi_5^0 + \phi_5^0 + \psi_5^0)(\mu) (1 - x_3) +
(\xi_5^- + \phi_5^- - \psi_5^-)(\mu) [2 x_1 x_2 - x_3 (1 - x_3)] \nnb \\
\ar(\xi_5^+ + \phi_5^+ + \psi_5^+)(\mu) [1 - x_3 - 2 (x_1^2 + x_2^2) ] \Big\}~, \nnb \\
T_5(x_i,\mu) \es 6 x_3 [\xi_5^0(\mu) + \xi_5^+(\mu) (1 - 2 x_3)]~, \nnb \\
T_6(x_i,\mu) \es 2 [\phi_6^0(\mu) + \frac{1}{2} (\phi_6^- - \phi_6^+)(\mu) 
(1 - 3 x_3)]~, \nnb \\
T_7(x_i,\mu) \es 6 x_3 \Big\{ (-\xi_4^0 + \phi_4^0 + \psi_4^0)(\mu) (1 - x_3) +
(-\xi_4^- + \phi_4^- - \psi_4^-)(\mu)[x_1^2 + x_2^2 - x_3 (1 - x_3)] \nnb \\
\ar (-\xi_4^+ + \phi_4^+ + \psi_4^+)(\mu) (1 - x_3 - 10 x_1 x_2) \Big\}~, \nnb \\
T_8(x_i,\mu) \es \frac{3}{2} \Big\{ (-\xi_5^0 + \phi_5^0 + \psi_5^0)(\mu) (1 - x_3) +
(-\xi_5^- + \phi_5^- - \psi_5^-)(\mu) [2 x_1 x_2 - x_3 (1 - x_3)] \nnb \\
\ar (-\xi_5^+ + \phi_5^+ + \psi_5^+)(\mu) [1 - x_3 - 2 (x_1^2 + x_2^2) ] \Big\}~, \nnb \\
S_1(x_i,\mu)  \es  6 x_3 (x_2 - x_1) \left[ (\xi_4^0 + \phi_4^0 +
\psi_4^0 + \xi_4^+ + \phi_4^+ + \psi_4^+)(\mu) + (\xi_4^- + \phi_4^-
- \psi_4^-)(\mu) (1-2 x_3) \right]~,
 \nnb \\
S_2(x_i,\mu)  \es  \frac{3}{2} (x_2 -x_1) [- (\psi_5^0 +
\phi_5^0 + \xi_5^0 )(\mu) + (\xi_5^- + \phi_5^- - \psi_5^- )(\mu) x_3 \nnb \\
\ar (\xi_5^+ + \phi_5^+ + \psi_5^+)(\mu) (1- 2 x_3) ]~, \nnb \\
P_1(x_i,\mu)  \es  6 x_3 (x_2-x_1) [ (\xi_4^0 - \phi_4^0 -
\psi_4^0 + \xi_4^+ - \phi_4^+ - \psi_4^+)(\mu) + (\xi_4^- - \phi_4^-
+ \psi_4^-)(\mu)(1-2 x_3)]~, \nnb \\
P_2(x_i,\mu)  \es  \frac32 (x_2 -x_1) [ (\psi_5^0 + \psi_5^0
- \xi_5^0)(\mu) + (\xi_5^- - \phi_5^- + \psi_5^-)(\mu) x_3 \nnb \\
\ar (\xi_5^+ - \phi_5^+ - \psi_5^+ )(\mu) (1- 2 x_3) ]~,
\end{eqnarray}
where $V_1$, $A_1$ and $T_1$ are leading twist--3; 
$S_1$, $P_1$, $V_2$, $V_3$, $A_2$, $A_3$, $T_2$,
$T_3$ and $T_7$ are twist--4; $S_2$,
$P_2$, $V_4$, $V_5$, $A_4$, $A_5$, $T_4$, $T_5$ and $T_8$ are
twist--5; and $V_6$, $A_6$ and $T_6$ are twist--6 DAs. 

\section{Non--perturbative parameters}
\setcounter{equation}{0}

The coefficients
$\phi_3^0$, $\phi_6^0$, $\phi_4^0$, $\phi_5^0$, $\xi_4^0$,
$\xi_5^0$, $\psi_4^0$, $\psi_5^0$, $\phi_3^-$, $\phi_3^+$,
$\phi_4^-$, $\phi_4^+$, $\psi_4^-$, $\psi_4^+$, $\xi_4^-$,
$\xi_4^+$, $\phi_5^-$, $\phi_5^+$, $\psi_5^-$, $\psi_5^+$,
$\xi_5^-$, $\xi_5^+$, $\phi_6^-$, $\phi_6^+ $ 
in the expansions of the DAs can be expressed in terms of eight
non--perturbative parameters $f_N$, $\lambda_1$, $\lambda_2$, $f_1^u$,
$f_1^d$, $f_2^d$, $A_1^u$, $V_1^d$ as follows (see \cite{R8117}):

{}For the leading confirmal spin,
\bea
\label{e81B1}
\phi_3^0 = \phi_6^0 = f_N \,,\hspace{0.15cm} 
&~&  
\phi_4^0 = \phi_5^0 = 
\frac{1}{2} \left(f_N + \lambda_1 \right) \,, 
\nnb \\
\xi_4^0 = \xi_5^0 = 
\frac{1}{6} \lambda_2\,,
&~&  
\psi_4^0  = \psi_5^0 =          
\frac12\left(f_N - \lambda_1 \right)\,,
\eea

{}for the next-to-leading spin, 
for twist-3:
\bea
\label{e81B2}
\phi_3^- = \frac{21}{2} f_N \, A_1^u, &~&
\phi_3^+ = \frac{7}{2}  f_N \, (1 - 3 V_1^d),
\eea

for twist-4:
\bea
\label{e81B3}
\phi_4^+ &=& \frac{1}{4} \left[ 
 f_N( 3 - 10  V_1^d ) + \lambda_1(3- 10 f_1^d) 
\right],
\nnb \\
\phi_4^- &=& - \frac{5}{4} \left[
f_N( 1 - 2 A_1^u ) - \lambda_1(1- 2 f_1^d -4 f_1^u)
\right],
\nnb \\
\psi_4^+ &=& - \frac{1}{4} \left[
f_N( 2\! +\! 5 A_1^u \!-\! 5 V_1^d) - \lambda_1 (2 \!-\! 5 f_1^d \!-\! 5 f_1^u) 
\right],
\nnb \\
\psi_4^- &=&  \frac{5}{4} \left[
f_N(2 - A_1^u - 3 V_1^d ) - \lambda_1(2- 7 f_1^d + f_1^u)
\right],
\nnb \\
\xi_4^+ &=& \frac{1}{16} \lambda_2 (4\!-\! 15 f_2^d)\,, ~
\xi_4^- = \frac{5}{16} \lambda_2(4\!-\! 15 f_2^d),
\eea

for twist-5: 
\bea
\label{e81B4}
\phi_5^+ &=& 
- \frac{5}{6} \left[
f_N( 3 + 4   V_1^d) - \lambda_1 (1 - 4 f_1^d )
\right],
\nnb \\
\phi_5^- &=& 
- \frac{5}{3} \left[
f_N( 1 - 2 A_1^u ) - \lambda_1(f_1^d - f_1^u) 
\right],
\nnb \\
\psi_5^+ &=& 
-\frac{5}{6} \left[
f_N(5\! + \!2 A_1^u \!-\!2 V_1^d) - \lambda_1 (1 \!-\! 2 f_1^d \!-\! 2 f_1^u)
\right],
\nnb \\
\psi_5^- &=& \phantom{-}
\frac{5}{3} \left[
f_N( 2 - A_1^u - 3 V_1^d) + \lambda_1 (f_1^d - f_1^u) 
\right],
\nnb \\
\xi_5^+ &=& \phantom{-}\frac{5}{36} \lambda_2 (2- 9 f_2^d) \,,
\; \; \; \;
\xi_5^- = - \frac{5}{4} \lambda_2 f_2^d\,,
\eea

and for twist-6: 
\bea
\label{e81B5}
\phi_6^+ &=& \frac{1}{2}\left[
f_N ( 1 - 4 V_1^d ) - \lambda_1  (1 - 2 f_1^d) 
\right],
 \\
\phi_6^- &=& \phantom{-}\frac{1}{2} \left[
f_N (1 +  4 A_1^u ) + \lambda_1 (1- 4 f_1^d - 2 f_1^u)
\right] \,.
\nnb
\eea
The values of the non--perturbative parameters $f_N$, $\lambda_1$, 
$\lambda_2$, $f_1^u$, $f_1^d$, $f_2^d$, $A_1^u$, $V_1^d$
can be found in \cite{R8105}
and \cite{R8117}.

\newpage

\section*{Figure captions}
{\bf Fig. (1)} The dependence of the scalar form factor of nucleon due to
the scalar current with quantum numbers of $\sigma$ meson, on Borel
parameter $M^2$ at $s_0=2~GeV^2$ for three different values of $Q^2$ and
$\beta$: $Q^2=3~GeV^2,Q^2=4~GeV^2,Q^2=5~GeV^2$ and $\beta=1,~\beta=3,~
\beta=5$. \\ \\
{\bf Fig. (2)} The same as in Fig. (1), but at $s_0=2.5~GeV^2$.\\ \\
{\bf Fig. (3)} The same as in Fig. (1), but for values of $\beta$:
$\beta=-1,~-3,~-5$. \\ \\
{\bf Fig. (4)} The same as in Fig. (2), but for values of $\beta$:
$\beta=-1,~\beta=-3,~\beta=-5$. \\ \\
{\bf Fig. (5)} The same as in Fig. (1), but for the scalar form factor
induced by the scalar current with quantum number of $a_0$ 
meson. \\ \\
{\bf Fig. (6)} The same as in Fig. (5), but at $s_0=2.5~GeV^2$.\\ \\
{\bf Fig. (7)} The same as in Fig. (5), but for values of $\beta$:
$\beta=-1,\beta=-3,~\beta=-5$. \\ \\
{\bf Fig. (8)} The same as in Fig. (7), but at $s_0=2.5~GeV^2$.\\ \\
{\bf Fig. (9)} The dependence of the scalar form factor induced by the
scalar current with $\sigma$ meson quantum numbers, on $Q^2$ at 
$M^2=1~GeV^2$ and $s_0=2~GeV^2$, at fixed values of 
$\beta=-1.4,~\beta=-3,~\beta=-5,~\beta=5$. \\ \\ 
{\bf Fig. (10)} The same as in Fig. (9), but at $s_0=2.5~GeV^2$.\\ \\
{\bf Fig. (11)} The dependence of the scalar form factor induced by the
scalar current with $a_0$ meson quantum numbers, on $Q^2$ at 
$M^2=1~GeV^2$ and $s_0=2~GeV^2$, at fixed values of 
$\beta=-1.4,~\beta=-3,~\beta=-5,~\beta=5$. \\ \\
{\bf Fig. (12)} The same as in Fig. (11), but at $s_0=2.5~GeV^2$.\\ \\

\newpage

\begin{figure}
\vskip 3. cm
    \includegraphics{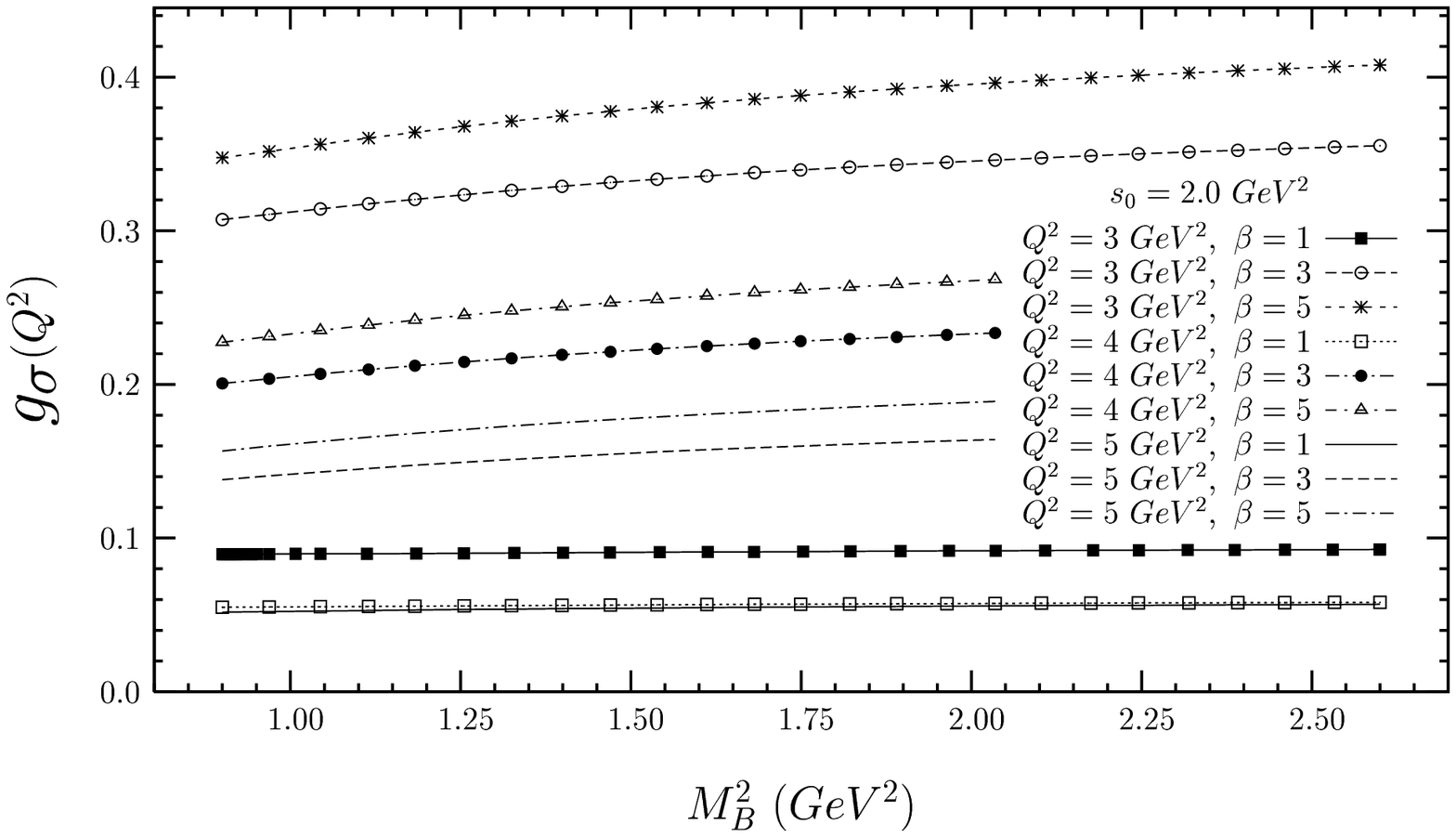}
\vskip 6.3cm
\caption{}
%\begin{center}
%{\bf Fig. 1--a}
%\end{center}
\end{figure}

\begin{figure}
\vskip 4.0 cm
    \includegraphics{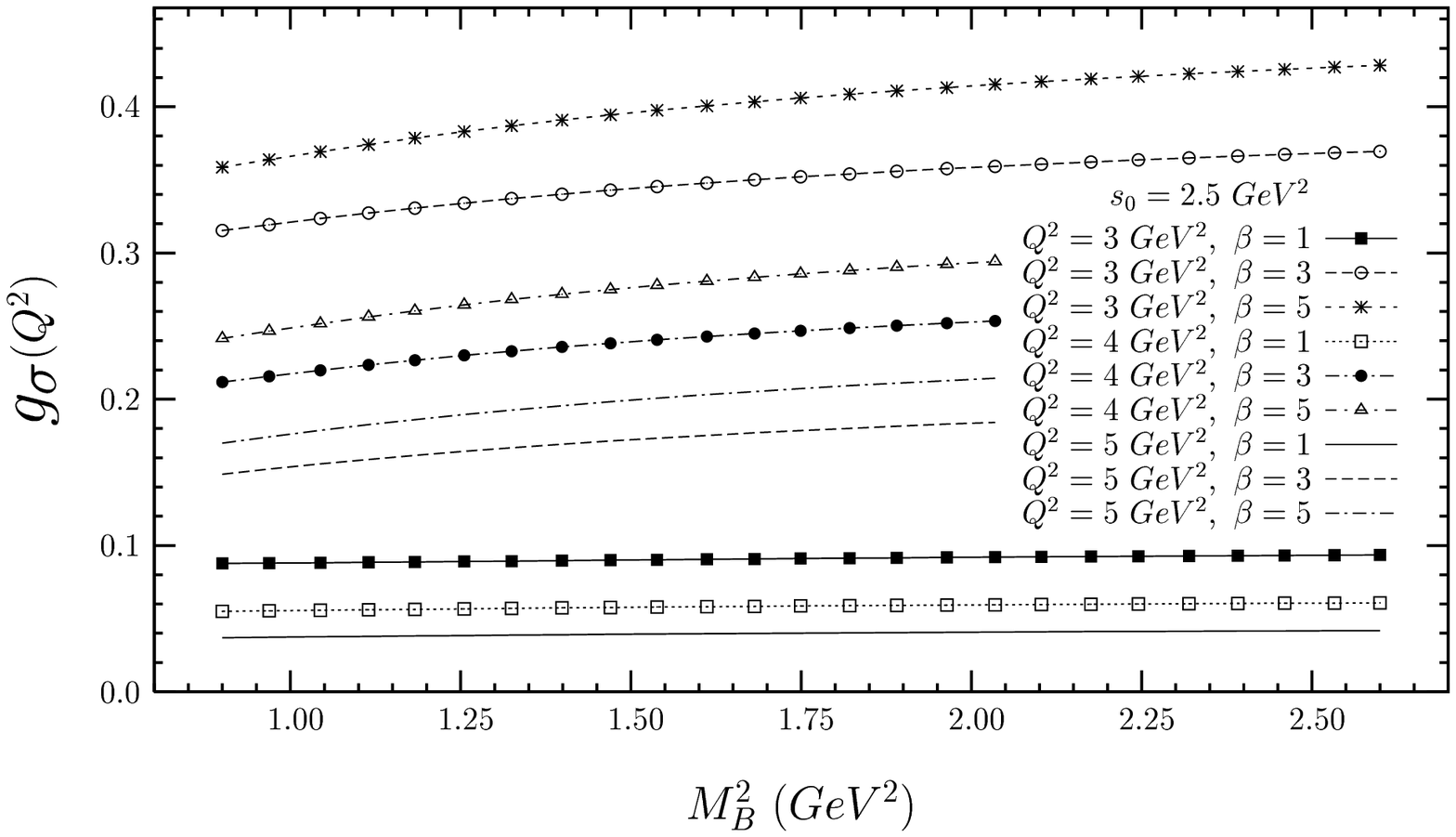}
\vskip 6.3 cm
\caption{}
%\begin{center}
%{\bf Fig. 1--b}
%\end{center}
\end{figure}

\newpage

\begin{figure}
\vskip 3. cm
    \includegraphics{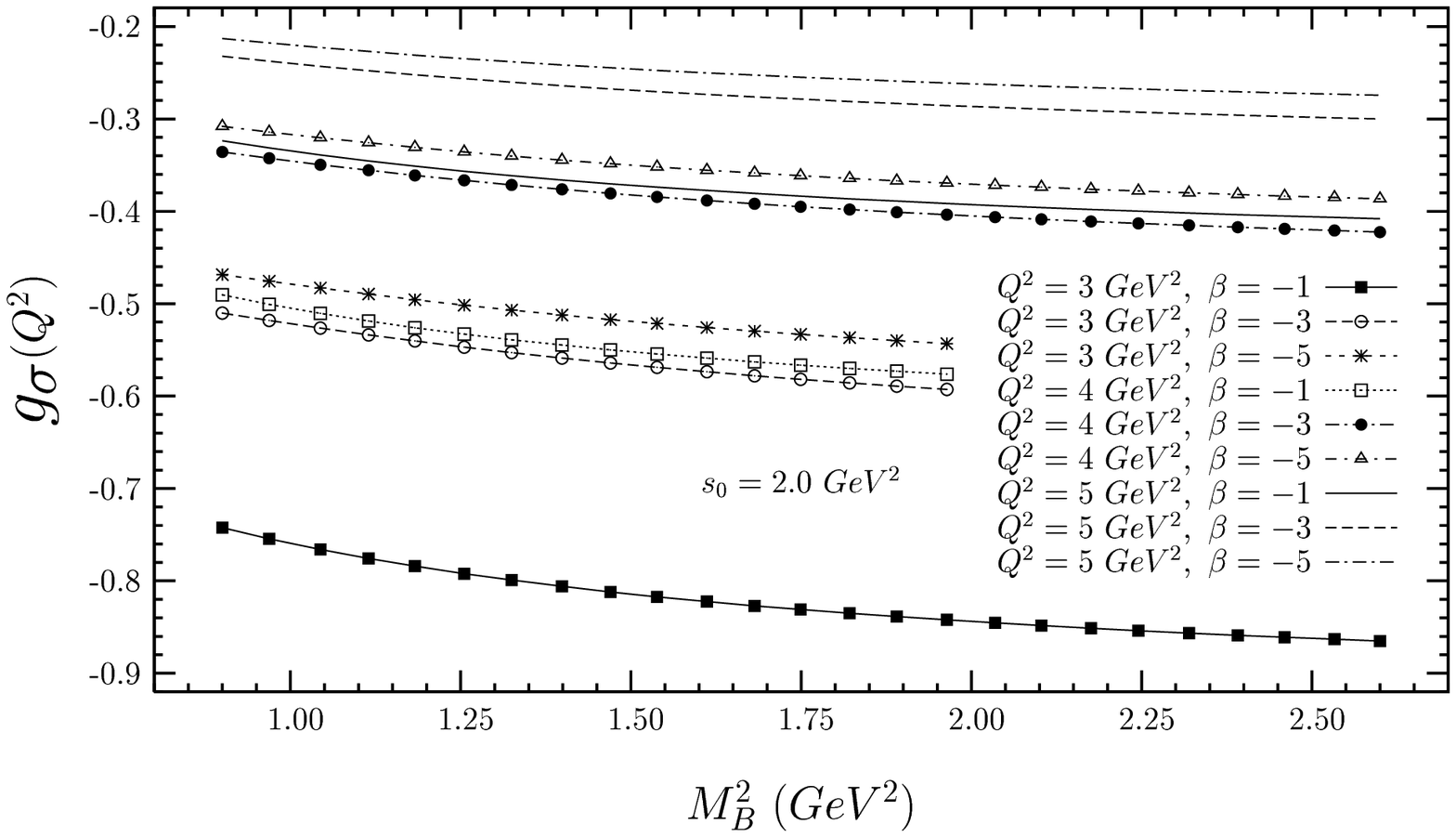}
\vskip 6.3cm
\caption{}
%\begin{center}
%{\bf Fig. 1--a}
%\end{center}
\end{figure}

\begin{figure}
\vskip 4.0 cm
    \includegraphics{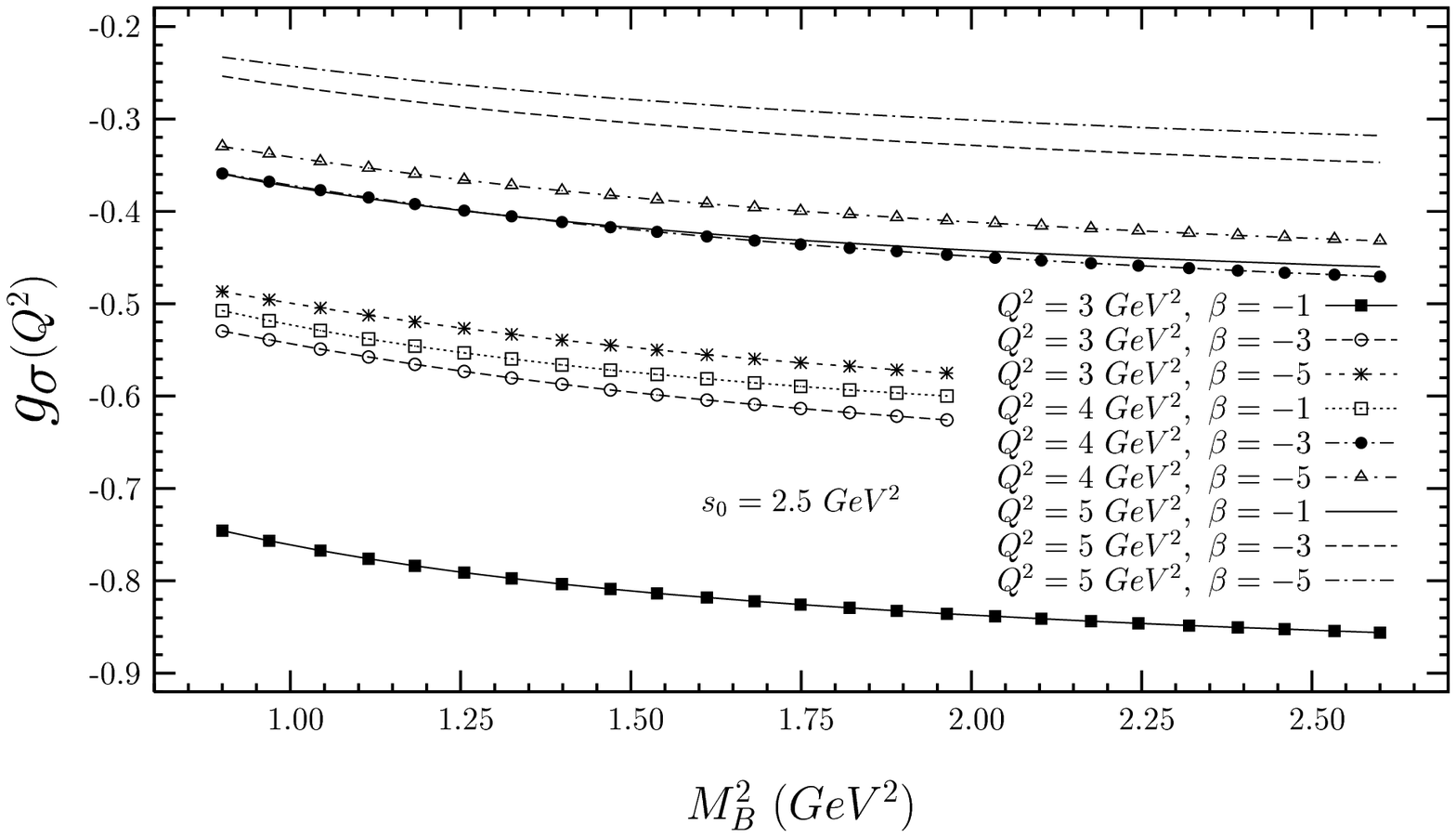}
\vskip 6.3 cm
\caption{}
%\begin{center}
%{\bf Fig. 1--b}
%\end{center}
\end{figure}

\newpage

\begin{figure}
\vskip 3. cm
    \includegraphics{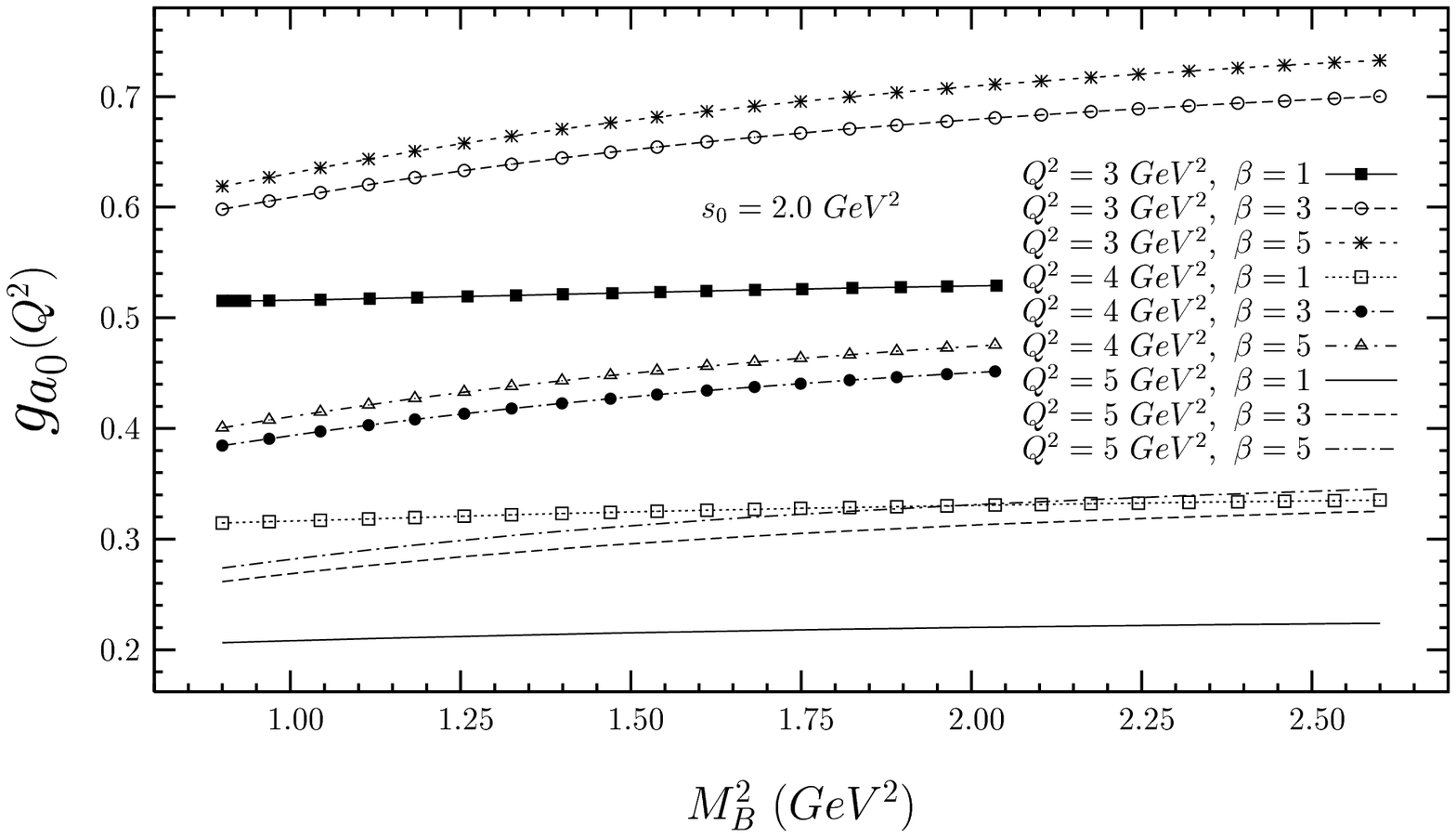}
\vskip 6.3cm
\caption{}
%\begin{center}
%{\bf Fig. 1--a}
%\end{center}
\end{figure}

\begin{figure}
\vskip 4.0 cm
    \includegraphics{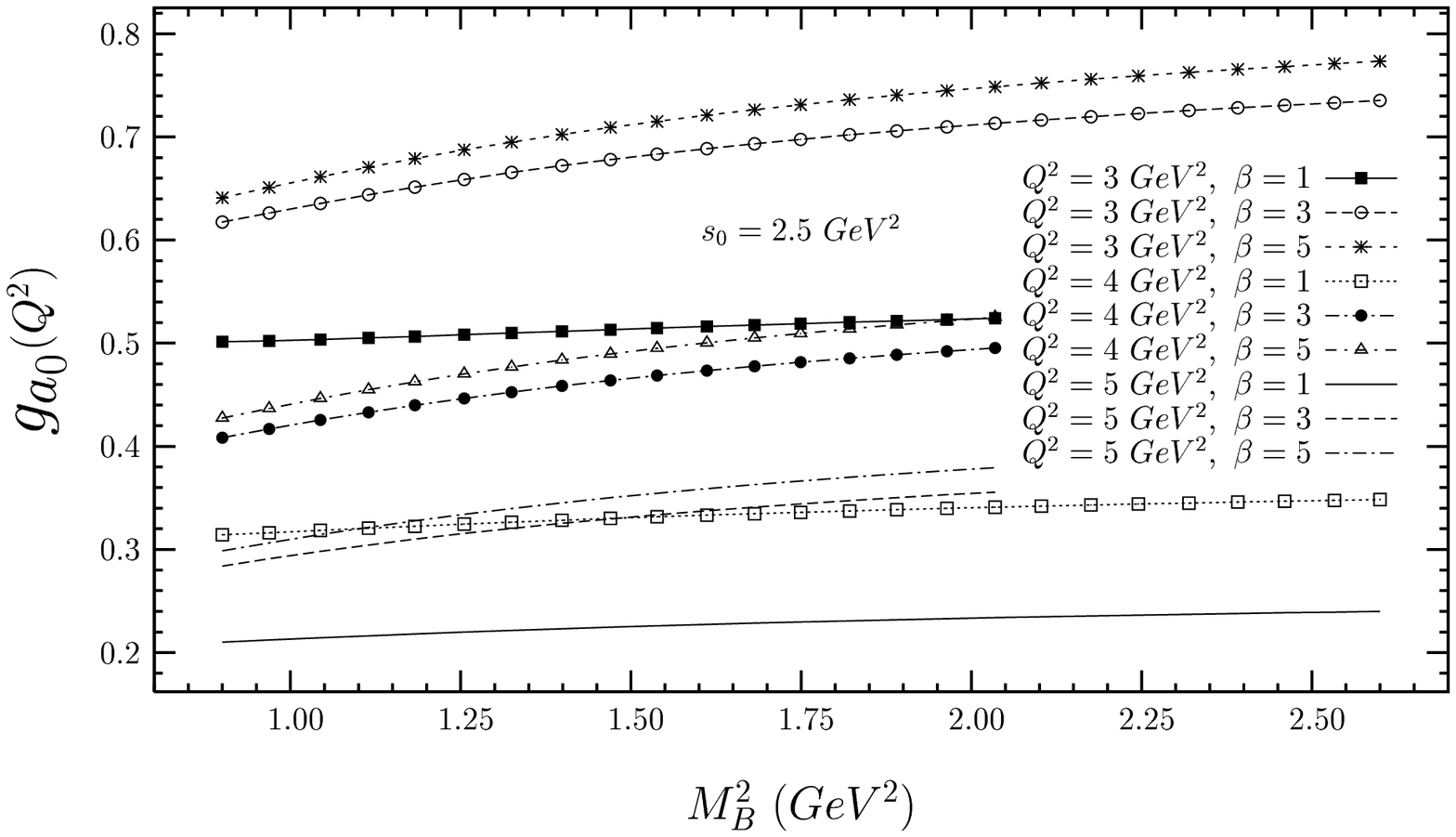}
\vskip 6.3 cm
\caption{}
%\begin{center}
%{\bf Fig. 1--b}
%\end{center}
\end{figure}

\newpage

\begin{figure}
\vskip 3. cm
    \includegraphics{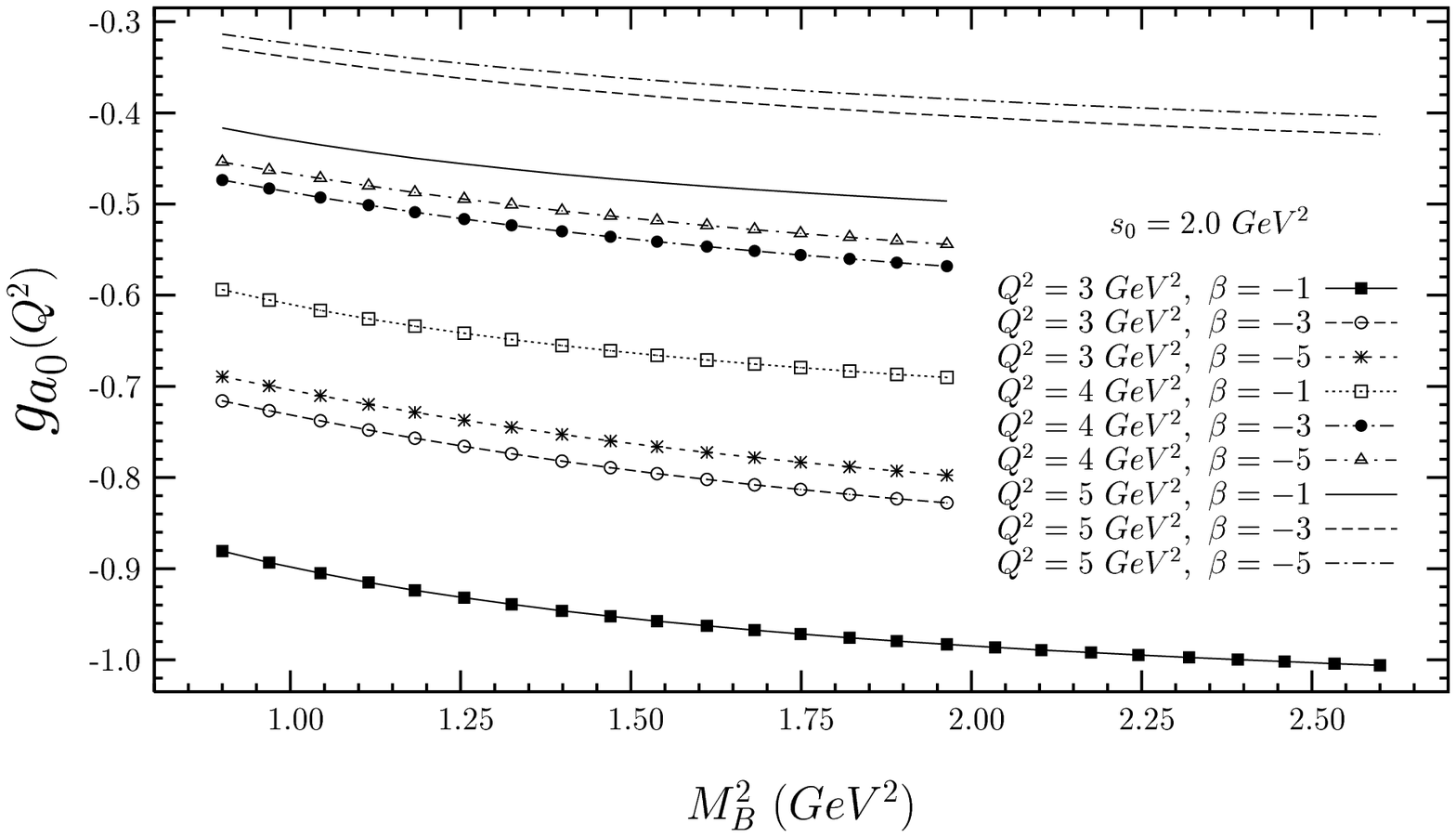}
\vskip 6.3cm
\caption{}
%\begin{center}
%{\bf Fig. 1--a}
%\end{center}
\end{figure}

\begin{figure}
\vskip 4.0 cm
    \includegraphics{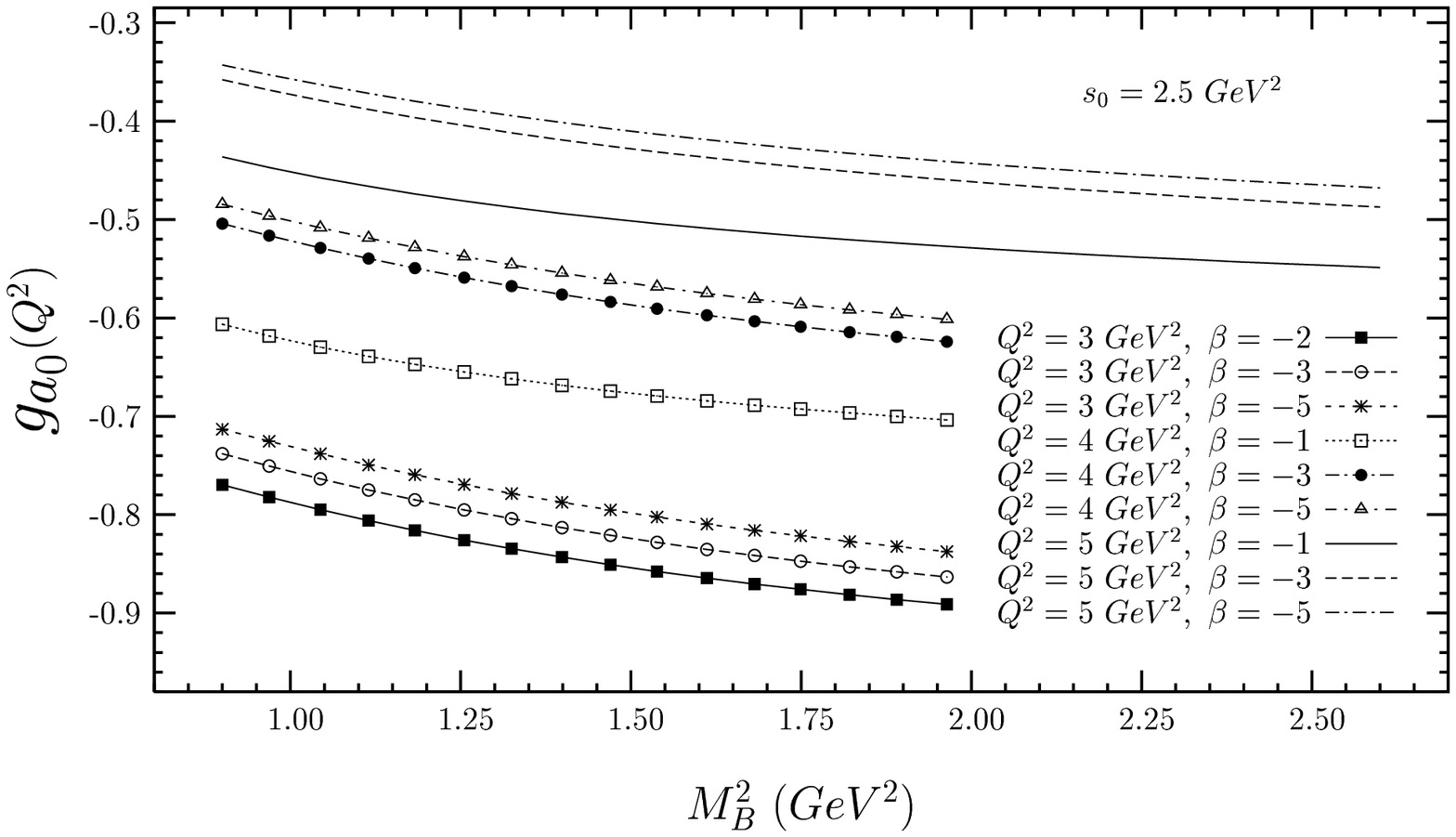}
\vskip 6.3 cm
\caption{}
%\begin{center}
%{\bf Fig. 1--b}
%\end{center}
\end{figure}

\newpage

\begin{figure}
\vskip 3. cm
    \includegraphics{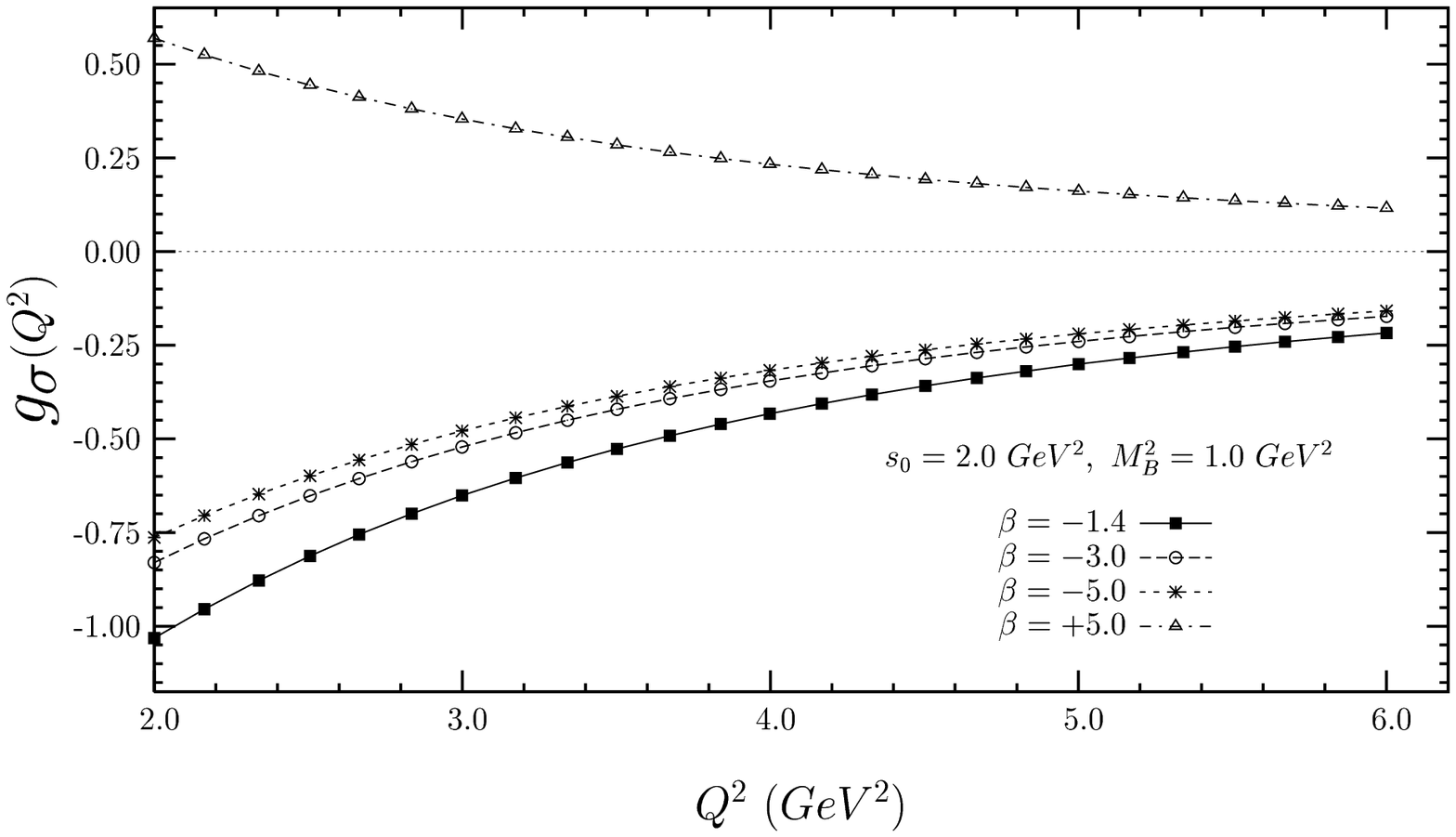}
\vskip 6.3cm
\caption{}
%\begin{center}
%{\bf Fig. 1--a}
%\end{center}
\end{figure}

\begin{figure}
\vskip 4.0 cm
    \includegraphics{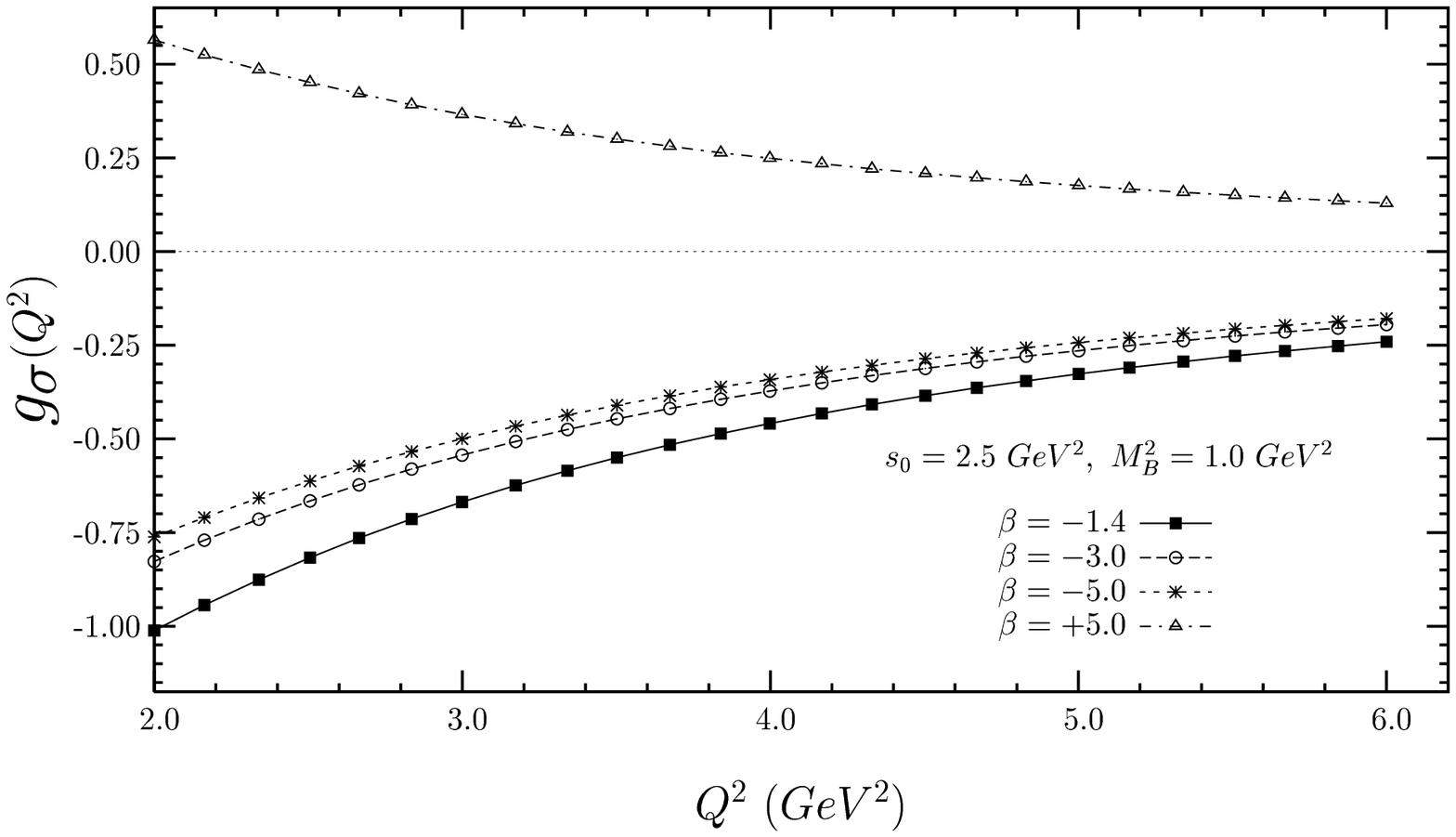}
\vskip 6.3 cm
\caption{}
%\begin{center}
%{\bf Fig. 1--b}
%\end{center}
\end{figure}

\newpage

\begin{figure}
\vskip 3. cm
    \includegraphics{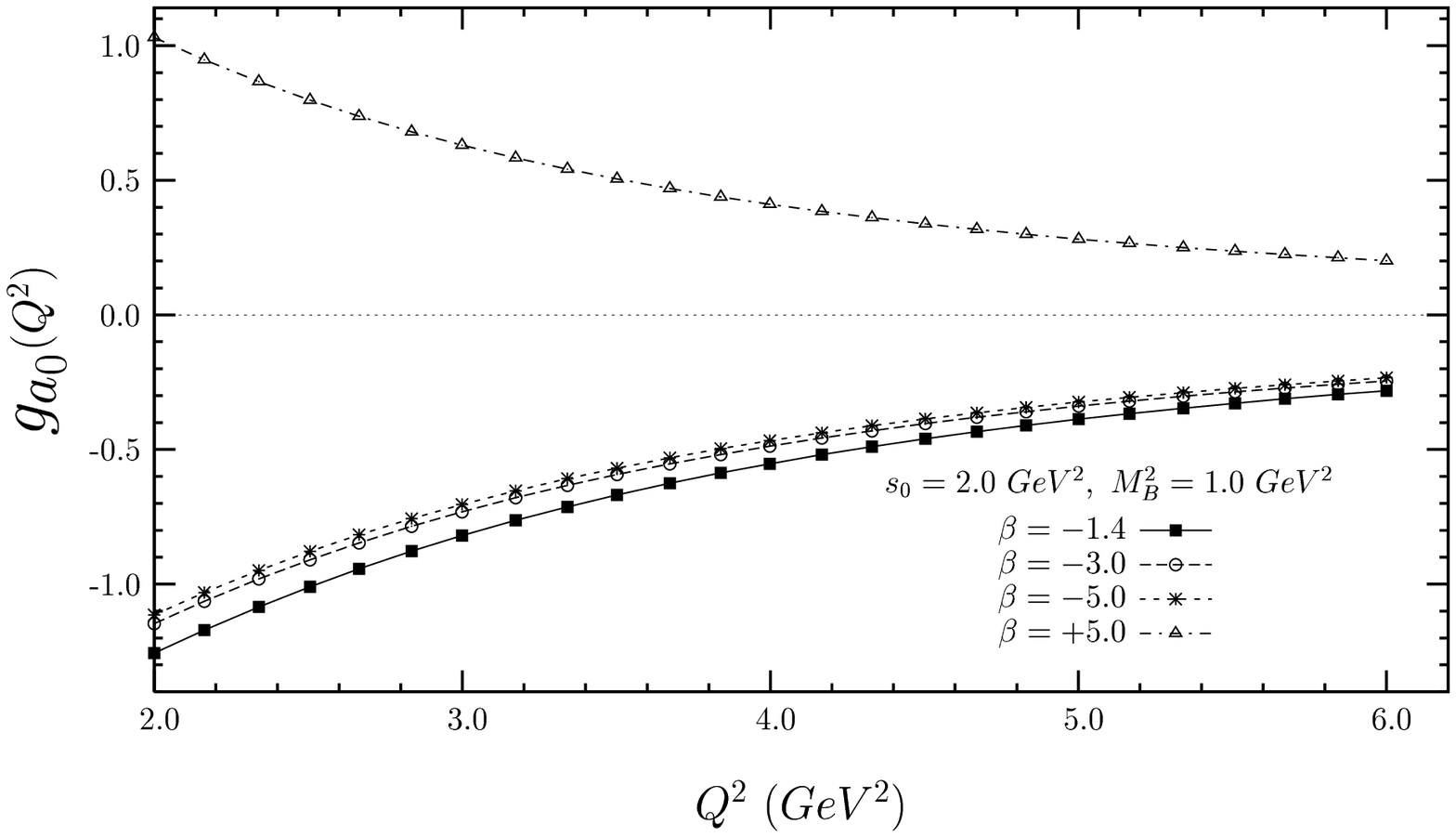}
\vskip 6.3cm
\caption{}
%\begin{center}
%{\bf Fig. 1--a}
%\end{center}
\end{figure}

\begin{figure}
\vskip 4.0 cm
    \includegraphics{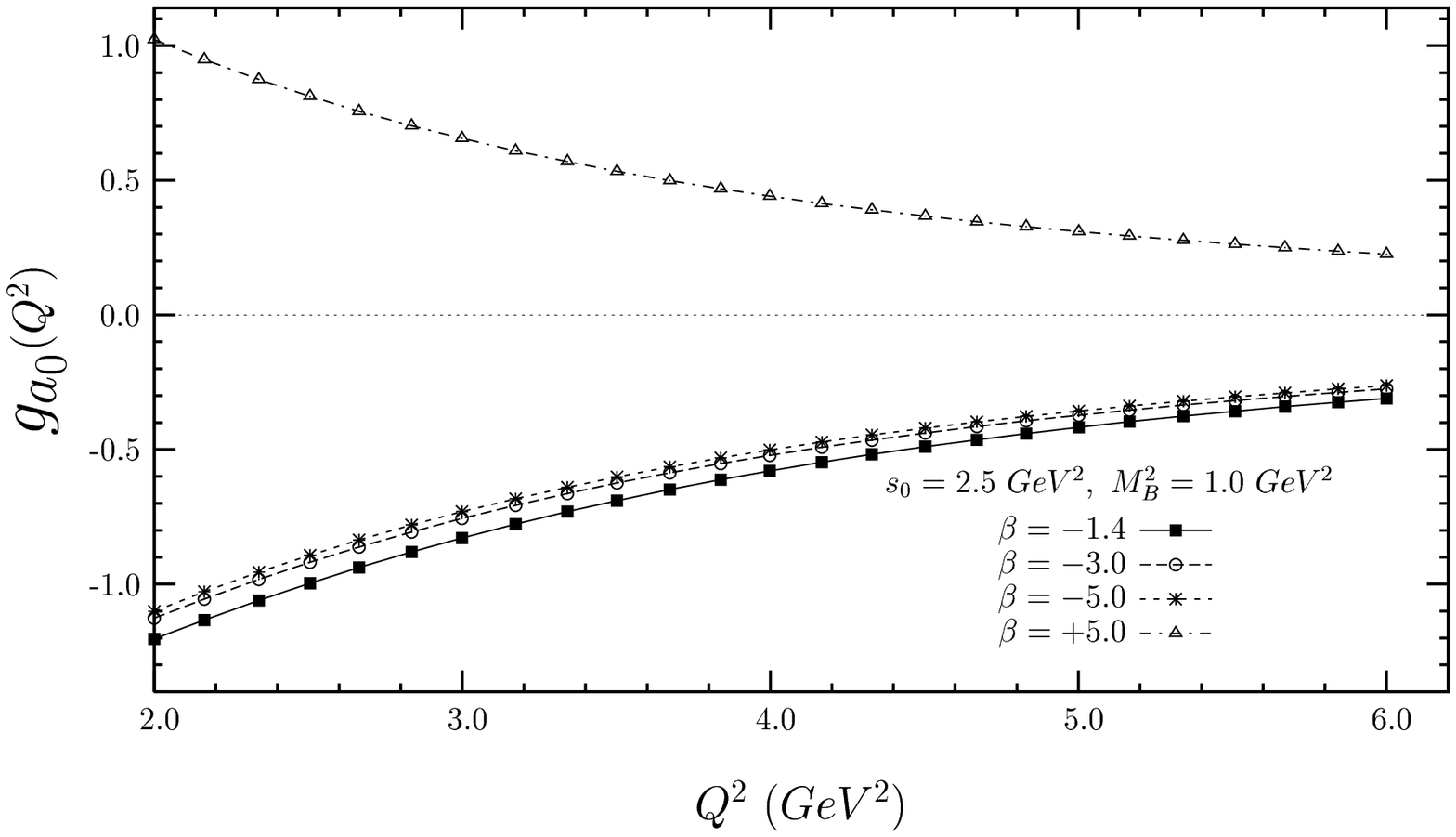}
\vskip 6.3 cm
\caption{}
%\begin{center}
%{\bf Fig. 1--b}
%\end{center}
\end{figure}

\end{document}